\documentclass[preprint,prd,onecolumn,preprintnumbers,superscriptaddress,showpacs,showkeys,nofootinbib]{revtex4-1}
\usepackage{graphicx,times}
\usepackage{amsmath,amsfonts,amssymb}
\usepackage{color}
\usepackage{psfrag}

\begin{document}

\title{Bound state formation in the $DDK$ system}
\author{A. Mart\'inez Torres}
\email{amartine@if.usp.br}
\affiliation{Instituto de F\'isica, Universidade de S\~ao Paulo, C.P. 66318, 05389-970 S\~ao 
Paulo, S\~ao Paulo, Brazil.}
\affiliation{School of Physics and Nuclear Energy Engineering \& Beijing Key Laboratory of Advanced Nuclear Materials and Physics, Beihang University, Beijing 100191, China}
\author{K. P. Khemchandani}
\email{kanchan.khemchandani@unifesp.br}
\affiliation{Universidade Federal de S\~ao Paulo, C.P. 01302-907, S\~ao Paulo, Brazil.}
\affiliation{School of Physics and Nuclear Energy Engineering \& Beijing Key Laboratory of Advanced Nuclear Materials and Physics, Beihang University, Beijing 100191, China}
\author{Li-Sheng Geng}
\email{lisheng.geng@buaa.edu.cn}
\affiliation{School of Physics and Nuclear Energy Engineering \& Beijing Key Laboratory of Advanced Nuclear Materials and Physics, Beihang University, Beijing 100191, China}
\affiliation{Beijing Advanced Innovation Center for Big Date-based Precision Medicine, Beihang University, Beijing100191, China.}
\date{\today}
\begin{abstract}
We study the $DDK$ system in a coupled channel approach, by including $DD_s\eta$ and $DD_s\pi$, and find that the dynamics involved in the system forms a bound state with isospin $1/2$ and mass $4140$ MeV
when one of the $DK$ pair is resonating in isospin 0, forming the $D^*_{s0}(2317)$. The state can be interpreted as a $DD^*_{s0}(2317)$ molecule like state with exotic quantum numbers: doubly charged, doubly charmed, and with single strangeness.
\end{abstract}


\maketitle

\date{\today}

\section{Introduction}
The existence of doubly charmed mesons and baryons is compatible with the present understanding of Quantum Chromodynamics and of the hadron structure~\cite{DeRujula:1975qlm,Gaillard:1974mw}. In this line, experimental and theoretical efforts, during the past years, have been dedicated
to the study of doubly charmed baryons, like $\Xi^+_{cc}$, $\Xi^{++}_{cc}$, $\Omega^+_{cc}$ (see, for example, Refs.~\cite{Vogt:1995tf,Mattson:2002vu,Albertus:2006ya,Wang:2010hs,Wang:2010vn,Karliner:2014gca,Can:2015exa,Aaij:2017ueg,Koshkarev:2016acq,Li:2017cfz,Richard,Blin,Liu:2018euh,Dias:2018qhp}), and doubly charmed mesons, like the $T_{cc}$ family and others (see, for example, Refs.~\cite{Zouzou:1986qh,Manohar:1992nd,Pepin:1996id,Janc:2004qn,Navarra:2007yw,Ding:2009vj,Vijande:2009kj,Molina:2010tx,Ikeda:2013vwa,Eichten:2017ffp,Hyodo:2017hue}). The existence of triple charm states has also been claimed~\cite{Chen:2017jjn}, and doubly charmed/bottom three body systems have been studied~\cite{Dias:2017miz,Ma:2017ery,Dias:2018iuy,Valderrama:2018knt}. However, in spite of all these efforts, the present available information about these states is still too preliminary to reach strong conclusions about their properties, and it still remains in the agenda of high energy physics to clarify the formation and nature of such states. At the same time, the situation is expected to improve since studies of hadrons with multicharm form a part of the present programs of several experimental facilities.

In this work, we embark on this odyssey and study the formation of bound states/resonances in a system of double charm and positive strangeness: the $DDK$ system. The motivation behind such a study is twofold: (1) The combination of the charm $D$ meson and the strange $K$ meson is known to give rise to an attractive interaction in the isospin 0, generating the $D^*_{s0}(2317)$ state~\cite{Aubert:2003fg,Besson:2003cp,Kolomeitsev:2003ac,vanBeveren:2003kd,Guo:2006fu,Gamermann:2006nm,Flynn:2007ki,Guo:2009ct,MartinezTorres:2011pr,Altenbuchinger:2013vwa,Torres:2014vna}. The addition of a $D$ meson to this system leads to a three-body system with attractive interactions in two subsystems. The $DD$ interaction is not attractive in nature and, thus, the dynamics involved does not generate a state. Still, the attraction in the two $DK$ pairs can dominate and form hadronic  bound states/resonances with a $DD^*_{s0}(2317)$ nature. (2) Such a possibility was recently studied in Ref.~\cite{SanchezSanchez:2017xtl}, treating the $DD^*_{s0}(2317)$ as an effective two body system and describing the $D-D^*_{s0}(2317)$ interaction through a Kaon exchange potential. Indeed, as a consequence, the generation of a bound state with a binding energy (with respect to the $D-D^*_{s0}(2317)$ threshold) of 15-50 MeV (depending on the kaon exchange potential considered) was predicted~\cite{SanchezSanchez:2017xtl}. 

Encouraged by these findings, in this work, we study the explicit three-body coupled channels $DDK$, $DD_s\pi$ and $DD_s\eta$ by solving the Faddeev equations~\cite{Faddeev:1960su} within the approach developed in Refs.~\cite{MartinezTorres:2007sr,Khemchandani:2008rk,MartinezTorres:2008gy,MartinezTorres:2008kh,MartinezTorres:2009xb,MartinezTorres:2009cw,MartinezTorres:2010zv,MartinezTorres:2011vh,Torres:2011jt}. For this, the input two-body $t$-matrices needed for the Faddeev equations are obtained by using effective Lagrangians and solving the Bethe-Salpeter equation in a coupled channel approach. As we show in this work, the dynamics involved in the three-body system leads to the formation of a bound state with spin-parity $J^P=0^-$, total isospin $1/2$, with a mass of 4140 MeV, i.e., around 90 MeV below the $DDK$ three-body threshold. This state is formed precisely when the coupled channel $DK$, $D_s\eta$ subsystems resonate and generate the $D^*_{s0}(2317)$.

\section{Formalism}
In the following subsection we first describe the method used to calculate the three-body scattering matrix for the coupled channel  $DDK$, $DD_s\pi$, $DD_s\eta$ system and, after it, in the next subsection, we give details on the approach used to determine the interaction between the different two-body subsystems.

\begin{figure}[h!]
\centering
\includegraphics[width=0.5\textwidth]{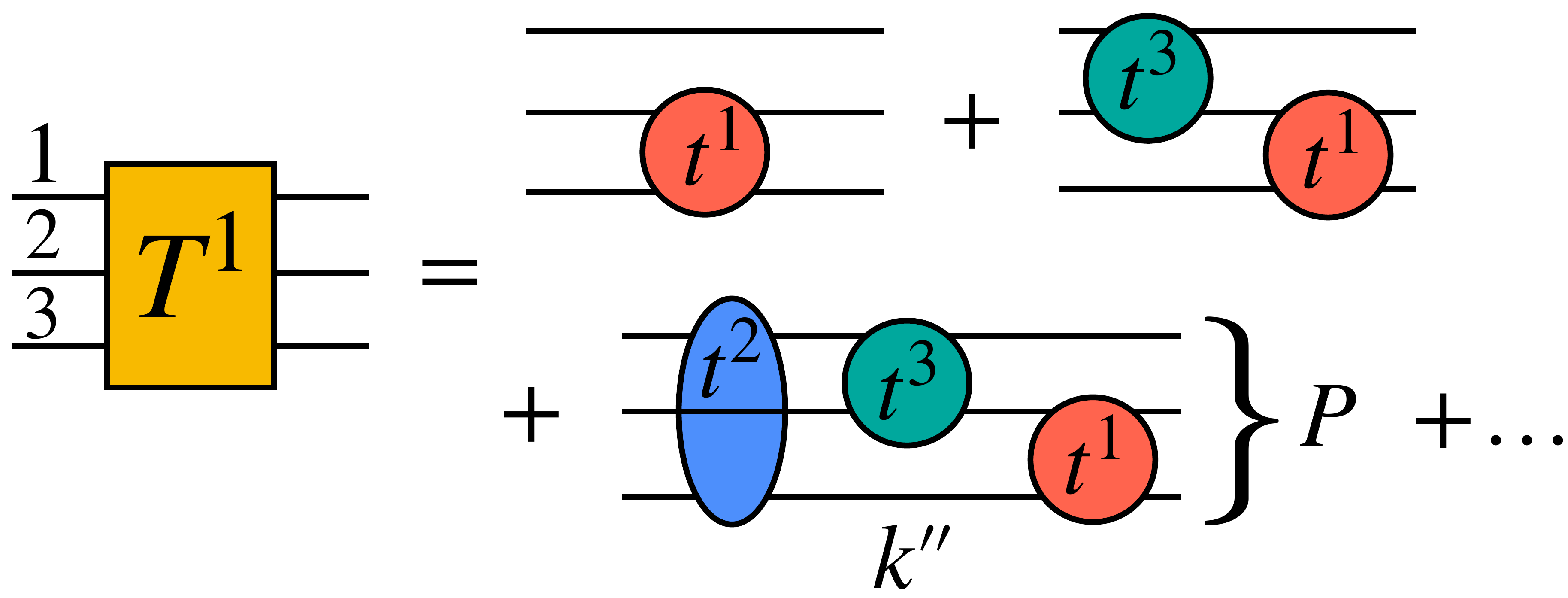}
\caption{Some diagrams contributing to the $T^1$ partition. The $P$ and $k^{\prime\prime}$ in the figure represent the four-momenta of the three-body system and of the indicated particle in the loop, respectively.}\label{T1fig}
\end{figure}

\subsection{The three-body problem}
The three-body scattering matrix $T$ describing the dynamics involved in the coupled channels can be written as a sum of three partitions~\cite{Faddeev:1960su}, $T^i$, $i=1,2,3$,
\begin{align}
T=\sum_{i=1}^3 T^i.\label{T}
\end{align}
Each of the partitions in Eq.~(\ref{T}) represents an infinite series of contributions to the scattering arising from Feynman diagrams where the $i$th particle is, by convention, a spectator in the right most interaction (see Fig.~\ref{T1fig}, for the case of the $T^1$ partition). Within the approach developed in Refs.~\cite{MartinezTorres:2007sr,Khemchandani:2008rk,MartinezTorres:2008gy,MartinezTorres:2008kh,MartinezTorres:2009xb,MartinezTorres:2009cw,MartinezTorres:2010zv,MartinezTorres:2011vh,Torres:2011jt}, each of the $T^i$ partitions can be expressed as
\begin{align}
T^i &=t^i\delta^3(\vec{k}^{\,\prime}_i-\vec{k}_i) + \sum_{j\neq i=1}^3T_R^{ij}, \quad i=1,2,3,\label{Ti}\\
T^{\,ij}_R &= t^ig^{ij}t^j+t^i\Big[G^{\,iji\,}T^{\,ji}_R+G^{\,ijk\,}T^{\,jk}_R\Big],  \label{TR}
\end{align}
where $\vec{k}_{i}$ ($\vec{k}^\prime_{i}$) corresponds to the initial (final) momentum of the particle $i$ and $t^{i}$ is the two-body $t$-matrix which describes the interaction of the $jk$ pair with $j \neq k\neq i=1,2,3$. In Eq.~(\ref{TR}), $g^{ij}$ represents the three-body Green's function of the system and the $G^{ijk}$ matrix is a loop function of three-particles. Their elements are defined as
\begin{align}
g^{ij} &(\vec{k}^\prime_i, \vec{k}_j)=\Bigg(\frac{N_{k}}{2E_k(\vec{k}^\prime_i+\vec{k}_j)}\Bigg)\nonumber\\
&\quad\times \frac{1}{\sqrt{s}-E_i
(\vec{k}^\prime_i)-E_j(\vec{k}_j)-E_k(\vec{k}^\prime_i+\vec{k}_j)+i\epsilon},\label{Green}
\end{align}
with $\sqrt{s}$ being the center of mass energy of the three-body system, $N_{k}=1$ for mesons, $E_{l}$, $l=1,2,3$, is the energy of the particle $l$, and
\begin{align}
G^{i\,j\,k}  =\int\frac{d^3 k^{\prime\prime}}{(2\pi)^3}\tilde{g}^{ij} \cdot F^{i\,j \,k}. \label{eq:Gfunc}
\end{align}
The elements of  $\tilde{g}^{ij}$ in Eq.~(\ref{eq:Gfunc}) are given by
\begin{align}
\tilde{g}^{ij}& (\vec{k}^{\prime \prime}, s_{lm}) = \frac{N_l}
{2E_l(\vec{k}^{\prime\prime})} \frac{N_m}{2E_m(\vec{k}^{\prime\prime})} \nonumber\\
&\quad\times
\frac{1}{\sqrt{s_{lm}}-E_l(\vec{k}^{\prime\prime})-E_m(\vec{k}^{\prime\prime})
+i\epsilon}, \quad i \ne l \ne m,
\label{eq:G} 
\end{align}
and the matrix $F^{i\,j\,k}$ in Eq.~(\ref{eq:Gfunc}), with explicit variable dependence, is given by 
\begin{align}
F^{i\,j\,k} &(\vec{k}^{\prime \prime},\vec{k}^\prime_j, \vec{k}_k,  s^{k^{\prime\prime}}_{ru})= t^{j}(s^{k^{\prime\prime}}_{ru}) g^{jk}(\vec{k}^{\prime\prime}, \vec{k}_k)\nonumber\\
&\quad\times\Big[g^{jk}(\vec{k}^\prime_j, \vec{k}_k) \Big]^{-1}
\Big[ t^{j} (s_{ru}) \Big]^{-1},  \quad j\ne r\ne u=1,2,3.\label{offac}
\end{align}
In Eq. (\ref{eq:G}), $\sqrt{s_{lm}}$ is the invariant mass of the $(lm)$ pair and can be calculated in terms of the external variables. The index $k^{\prime\prime}$ on the invariant mass $s^{k^{\prime\prime}}_{ru}$ of Eq.~(\ref{offac}) indicates its dependence on the loop variable
\begin{align}\label{offshell_IM}
s^{k^{\prime\prime}}_{ru}=(P-k^{\prime\prime})^2,
\end{align}
which, in turn, indicates the off-shell dependence of the amplitudes present in the loop. For example, the $t^3$-matrix, in the third diagram in Fig.~\ref{T1fig}, represents the interaction of particles 1 and 2 (thus, the index 3 indicates the label of the spectator particle) in the loop. This amplitude is calculated as a function of $s^{k^{\prime\prime}}_{12}$ using Eq.~(\ref{offshell_IM}) (see Refs. \cite{MartinezTorres:2007sr,Khemchandani:2008rk,MartinezTorres:2008gy,MartinezTorres:2008kh,MartinezTorres:2009xb,MartinezTorres:2009cw,MartinezTorres:2010zv,MartinezTorres:2011vh,Torres:2011jt} for more details).  

The $T^{ij}_R$ partitions obtained from Eq.~(\ref{TR}) are functions of two variables: the total three-body energy of the system, $\sqrt s$, and the 
invariant mass of the particles 2 and 3, $\sqrt{s_{23}}$. The other invariant masses, $\sqrt{s_{12}}$ and $\sqrt{s_{31}}$ can be obtained in terms of  $\sqrt s$ and $\sqrt{s_{23}}$, as shown in Refs. \cite{MartinezTorres:2007sr,Khemchandani:2008rk}. We study the behavior of the modulus square of the three-body $T$-matrix as a function of the total energy and the invariant mass of a subsystem. We do this by making three-dimensional plots and a peak appearing in such plots is interpreted as resonances/bound states linked to the three-body dynamics involved in the system under study. Since the first term in Eq.~(\ref{Ti}) can not give rise to any three-body state, we study the properties of the $T_R$ matrix defined as
\begin{align}
T_R\equiv \sum_{i=1}^3\sum_{j\neq i=1}^{3} T^{(ij)}_R.
\end{align}

We work in the charge basis taking into account the following channels: $D^+D^0 K^+$, $D^+D^+K^0$, $D^+D^+_s\pi^0$, $D^+D^+_s\eta$, $D^0 D^+ K^+$, $D^0D^+_s\pi^+$. To identify the peaks found in the three-body $T$-matrix with physical states, an isospin projection of the amplitudes is required. To do this, we use a basis in which the states are labeled in terms of the total isospin $I$ of the three-body system and the isospin of one of the two-body subsystems, which in the present case is taken as the isospin of the $DK$ subsystem labelled as particles 2 and 3, $I_{23}$, and evaluate the transition amplitude $\langle I,I_{23}|T_R|I,I_{23}\rangle$. The isospin  $I_{23}$ can be 0 or 1, thus, the total isospin $I$ can be 1/2 or 3/2. 

\subsection{The two-body scattering matrix}
In Refs.~\cite{Kolomeitsev:2003ac,Guo:2006fu,Gamermann:2006nm,Guo:2017jvc} it was shown that the interaction of the $DK$ and $\eta D_s$ coupled channel system in the isospin 0 configuration generates $D^*_{s 0} (2317)$.  The starting point in the latter works consists of using Lagrangians based on symmetries like chiral and heavy quark~\cite{Burdman:1992gh,Yan:1992gz}, relevant to such systems, to obtain the lowest order amplitude, $V$, describing the transition between the different coupled channels and unitarize the amplitudes. The unitarization is achieved by using $V$ as kernel in the Bethe-Salpeter equation, obtaining in this way the scattering matrix $t$ for the coupled channel system. This is done by solving the Bethe-Salpeter equation in its on-shell factorization form~\cite{Oller:1997ti,Jido:2003cb,Oller:1998hw}, 
\begin{align}
t=(1-V\mathcal{G})^{-1}V. \label{BS}
\end{align}
The $\mathcal{G}$ in Eq.~(\ref{BS})  represents the loop function of two hadrons, which has to be regularized either with a cut-off or dimensional regularization.

In case of the $DDK$ system and coupled channels, the resolution of Eq.~(\ref{TR}) requires the two-body $t$-matrices related to the $DK$ and $DD$ subsystems, and their respective coupled channels. To calculate the scattering matrix of the $DK$ system, which is formed by a heavy meson $H$ (the $D$ meson) and a light pseudoscalar $P$ (the Kaon), we follow closely the approach developed in Refs.~\cite{Guo:2006fu,Guo:2017jvc}. In these works, the leading order Lagrangian describing the $HP$ interaction is given by the kinetic and mass term of the heavy mesons (chiraly coupled to pions, since both chiral and heavy quark symmetries should be relevant to the problem), 
\begin{align}
\mathcal{L}=D_{\mu}HD^{\mu}H^{\dagger}-\mathring{M}^{2}_{H} H H^{\dagger}\label{lheavy},
\end{align}
with $H=\left(\begin{array}{ccc}D^{0}&D^{+}&D^{+}_{s}\end{array}\right)$ collecting the heavy mesons, whose mass in the chiral limit is $\mathring{M}_{H}$, $P$ is given by
\begin{align}
P=\left(\begin{array}{ccc}\frac{1}{\sqrt{2}}\pi^0+\frac{1}{\sqrt{6}}\eta & \pi^+ & {K}^+ \\ \pi^- & -\frac{1}{\sqrt{2}}\pi^0+\frac{1}{\sqrt{6}}\eta & {K}^0 \\{K}^- & \bar{K}^{0} & -\frac{2}{\sqrt{6}}\eta\end{array}\right),\label{phi}
\end{align}
and $D_{\mu}$ is the covariant derivative~\cite{Yan:1992gz}
\begin{align}
D_{\mu}H^{\dagger}&=(\partial_{\mu}+\Gamma_{\mu})H^{\dagger},\quad
D_{\mu}H=H(\overleftarrow{\partial}_{\mu}+\Gamma^{\dagger}_{\mu}),\\
\Gamma_{\mu}&=\frac{1}{2}(u^{\dagger}\partial_{\mu}u+u\partial_{\mu}u^{\dagger}),\quad
u^{2}=e^{i\sqrt{2}P/f}.\nonumber
\end{align}
For the case in which we are interested, i.e.,  $H P\to HP$, Eq.~(\ref{lheavy}) reduces to the following Lagrangian
\begin{align}
\mathcal{L}_{HP}=\frac{1}{4f^{2}}\left\{\partial^{\mu}H[P,\partial_{\mu}P]H^{\dagger}-H[P,\partial_{\mu}P]\partial^{\mu}H^{\dagger}\right\},\label{heavy}
\end{align}
and the lowest order amplitude obtained from it, in terms of the Mandelstam variables, reads as
\begin{align}
V_{ij}=-\frac{C_{ij}}{4f^{2}}(s-u).\label{pot}
\end{align}

In Eq.~(\ref{pot}) the $i$ and $j$ subindices represent the initial and final channels, respectively, and the $C_{ij}$ coefficients can be found in Refs.~\cite{Guo:2006fu,Guo:2017jvc}. This amplitude $V_{ij}$ is further projected on s-wave.

As in Refs.~\cite{Guo:2006fu,Guo:2017jvc}, we consider $D^0K^+$, $D^+ K^0$, $D^+_{s}\eta$ and $D^+_{s}\pi^0$ as coupled channels and regularize the loop function of Eq.~(\ref{BS}) using dimensional regularization with a scale $\mu=1000$ MeV and subtraction constant $a(\mu)=-1.846$. The resolution of Eq.~(\ref{BS}) for this system generates a pole, which is below the threshold of the $DK$ channel, with total isospin $0$, at 2318 MeV, and which can be associated with $D_{s^* 0}(2317)$.

The $DD$ and $DD_s$ interactions have not been studied, so far, but we can obtain the relevant amplitudes by following the procedure adopted in Ref.~\cite{Sakai:2017avl} to study the $BD$ system. In Refs.~\cite{Sakai:2017avl,Molina:2009ct,Ozpineci:2013qza,Dias:2014pva} the approach based on the hidden local symmetry~\cite{Bando:1984ej,Bando:1987br}, where the interactions proceed through vector meson exchange, has been extended to the sectors of charm and beauty.  In such an approach, which has been shown to be compatible with heavy quark symmetry~\cite{Sakai:2017avl,Molina:2009ct,Ozpineci:2013qza,Dias:2014pva}, the vector-pseudoscalar-pseudoscalar Lagrangian is written as
\begin{align}
\mathcal{L}_{VPP}&=-ig\langle V^\mu[\phi,\partial_\mu \phi]\rangle,\quad g=\frac{M_V}{2f},\label{LVPP}
\end{align}
where $M_V$ is the mass of the exchanged vector meson, $f=f_D=165$ MeV and
\begin{widetext}
\begin{align}
\phi=\left(\begin{array}{cccc}\frac{1}{\sqrt{2}}\pi^0+\frac{1}{\sqrt{3}}\eta +\frac{\eta^\prime}{\sqrt{6}}& \pi^+ &K^{+}&\bar{D}^{0}\\\pi^-&-\frac{1}{\sqrt{2}}\pi^0+\frac{1}{\sqrt{3}}\eta+\frac{\eta^\prime}{\sqrt{6}}&K^{0}&D^{-}\\K^{-}&\bar{K}^{0}& -\frac{1}{\sqrt{3}}\eta+\sqrt{\frac{2}{3}}\eta&D^{-}_s\\D^{0}&D^{+}&D^{+}_s&\eta_c\end{array}\right),\label{phi2}
\end{align}
\end{widetext}
\begin{align}
V^\mu=\left(\begin{array}{cccc}\frac{\omega}{\sqrt{2}}+\frac{\rho^0}{\sqrt{2}}&\rho^+&K^{*+}&\bar{D}^{*0}\\\rho^-&\frac{\omega}{\sqrt{2}}-\frac{\rho^0}{\sqrt{2}}&K^{*0}&D^{*-}\\K^{*-}&\bar{K}^{*0}&\phi&K^{*-}_s\\D^{*0}&D^{*+}&D^{*+}_s&J/\psi\end{array}\right).
\end{align}
Using Eq.~(\ref{LVPP}), the process $D_1D_2\to D^\prime_1D^\prime_2$, where $D_i$ ($D^\prime_i$), $i=1,2$, represent an initial (final) $D$ meson, gets contributions from the exchange of vector mesons in the $t-$ and $u-$channels. These contributions are given by
\begin{align}
V^\text{$t$-ch}_{ij}&=-g^2 \Big[(2s+t-m^2_{1i}-m^2_{2i}-m^2_{1j}-m^2_{2j})C^\text{$t$-ch}_1\nonumber\\
&\quad-(m^2_{1i}-m^2_{1j})(m^2_{2j}-m^2_{2i})C^\text{$t$-ch}_2\Big],\nonumber\\
V^\text{$u$-ch}_{ij}&=-g^2 \Big[(2s+u-m^2_{1i}-m^2_{2i}-m^2_{1j}-m^2_{2j})C^\text{$u$-ch}_1\nonumber\\
&\quad-(m^2_{1i}-m^2_{2j})(m^2_{1j}-m^2_{2i})C^\text{$u$-ch}_2\Big],\label{Vtu}
\end{align}
where
\begin{align}
C_1^\text{$t$-ch}&=\sum_k \frac{A^\text{$t$-ch}_{k}}{t-m^2_{Vk}+i\epsilon},\nonumber\\
C_2^\text{$t$-ch}&=\sum_k \frac{A^\text{$t$-ch}_{k}}{m^2_{Vk}(t-m^2_{Vk}+i\epsilon)},\nonumber\\
C_1^\text{$u$-ch}&=\sum_k \frac{B^\text{$u$-ch}_{k}}{u-m^2_{Vk}+i\epsilon},\label{Cs}\\
C_2^\text{$u$-ch}&=\sum_k \frac{B^\text{$u$-ch}_{k}}{m^2_{Vk}(u-m^2_{Vk}+i\epsilon)},\nonumber
\end{align}
with the index $k$ indicating the exchanged vector meson ($\rho$, $\omega$, $J/\psi$) of mass $m_{Vk}$. In the energy region studied for the three-body system, the invariant mass of the $DD$ system is near the $DD$ threshold, thus, when exchanging vector mesons in the $t-$ and $u-$channels, we can write $1/(t-m^2_{Vk}+i\epsilon)$ and $1/(u-m^2_{Vk}+i\epsilon)$ $\sim$ $-1/m^2_{Vk}$. The coefficients $A^\text{$t$-ch}_{k}$ ($B^\text{$u$-ch}_{k}$) of Eq.~(\ref{Cs}) are related to the two vertices involved in the t-channel (u-channel) exchange of vector mesons and are obtained from the Lagrangian in Eq.~(\ref{LVPP}) (see the tables in the Appendix~\ref{coef} for their specific values). The potentials in Eq.~(\ref{Vtu}) are summed and projected on s-wave. The result found is used to solve Eq.~(\ref{BS}) with the loop function $G$ regularized within the dimensional regularization scheme (in this case, we consider the regularization scale $\mu=1500$ MeV and the subtraction constant $a(\mu)=-1.3$ as in Refs.~\cite{Gamermann:2006nm,Gamermann:2007fi}). As we shall discuss in the next section, we vary this parameter to study the stability of the results.

\subsection{Off-shell contributions and three-body forces}
It is interesting to notice that Eq.~(\ref{TR}) is a set of six coupled matrix equations which are solved by using the on-shell part of the two-body $t$-matrices. It was shown in Refs.~\cite{MartinezTorres:2007sr,Khemchandani:2008rk,MartinezTorres:2008gy,Torres:2011jt} that this is due to the finding of a cancellation between the contribution of the off-shell parts of the two-body $t$-matrices to the three-body Faddeev amplitudes and a contact term with same topology whose origin is in the same Lagrangian which is used to describe the two-body interactions. For the case of a system formed by two pseudoscalar mesons and a baryon or two pseudoscalar mesons and a vector meson with $S$-wave interactions it was found that this cancellation was exact in the flavor SU(3) limit. In case of a system of three light pseudoscalar mesons in $S$-wave, such cancellation was found to be exact in the chiral limit (in this case, two more diagrams were taken into account, involving one and five meson intermediate states~\cite{Torres:2011jt}). In a realistic case, off such limits, in all these systems, the sum of the contributions related to the off-shell parts and the three-body contact term has always been estimated to be smaller than 5-7\% of the total on-shell contribution. Thus, only the on-shell part of the two-body $t$ matrices has been found to be significant and, for the purpose of investigating the formation of a three-body state and its properties,  the contribution coming from the off-shell parts of the two-body $t$ matrices used, and the one related to the three-body contact term obtained from the Lagrangian (and the one from the diagrams with one and five meson intermediate states in case of a three light pseudoscalar system), have together been neglected. In the present system, formed by two heavy pseudoscalars and a light one interacting in $S$-wave, the situation is slightly different: the Lagrangian used in Eq.~(\ref{lheavy}) to determine the two-body amplitudes does not generate a three-body contact term since it can not be expanded up to any desired number of heavy hadron fields. In such a case, and having in mind the findings of Refs.~\cite{MartinezTorres:2007sr,Khemchandani:2008rk,MartinezTorres:2008gy,Torres:2011jt}, we might expect a cancellation between the contributions obtained from the off-shell part of the two-body amplitudes to the Faddeev equations under some limit and, when being off such limit, such contributions should be much smaller than those obtained from the on-shell part of the two-body amplitudes as in Refs.~\cite{MartinezTorres:2007sr,Khemchandani:2008rk,MartinezTorres:2008gy,Torres:2011jt}. We show in Appendix~\ref{cancell} that this is precisely the situation here: similarly to the case of a three light pseudoscalar system, an exact cancellation occurs between the off-shell contributions of the two-body amplitudes to the different three-body diagrams in the chiral limit. In a realistic situation the contribution arising from the off-shell parts is found to be about $1\%$ of the total on-shell contribution and we, thus, neglect it. 

One might also wonder what would happen with such a cancellation if a different model, not based on heavy quark symmetry, and which could generate a three-body contact term, is used to determine the two-body amplitudes required to solve the Faddeev equations. To answer such a question, we have also considered the model of Ref.~\cite{Gamermann:2006nm}, which is based on the $SU(4)$ symmetry and which describes well the properties of the $D^*_{s0}(2317)$ state as a $DK$ bound state. As shown in Appendix~\ref{cancell}, in such a model, there exits an explicit three-body contact term as in case of Refs.~\cite{MartinezTorres:2007sr,Khemchandani:2008rk,MartinezTorres:2008gy,Torres:2011jt}. We find that, in this case too, an exact cancellation exists in the chiral limit and away from the limit the total contribution of the different sources of three-body contact terms remains small. It is also interesting to mention that, numerically, the contributions arising from the off-shell as well as the on-shell parts of the two-body amplitudes, to the three-body diagrams, when calculated with the heavy quark model of Refs.~\cite{Guo:2006fu,Guo:2017jvc} is almost identical to those coming from the two-body amplitudes and the three-body contact term determined with the model of Ref.~\cite{Gamermann:2006nm}, which is based on SU(4).

\section{Results}
\begin{figure}[h!]
\centering
\includegraphics[width=0.5\textwidth]{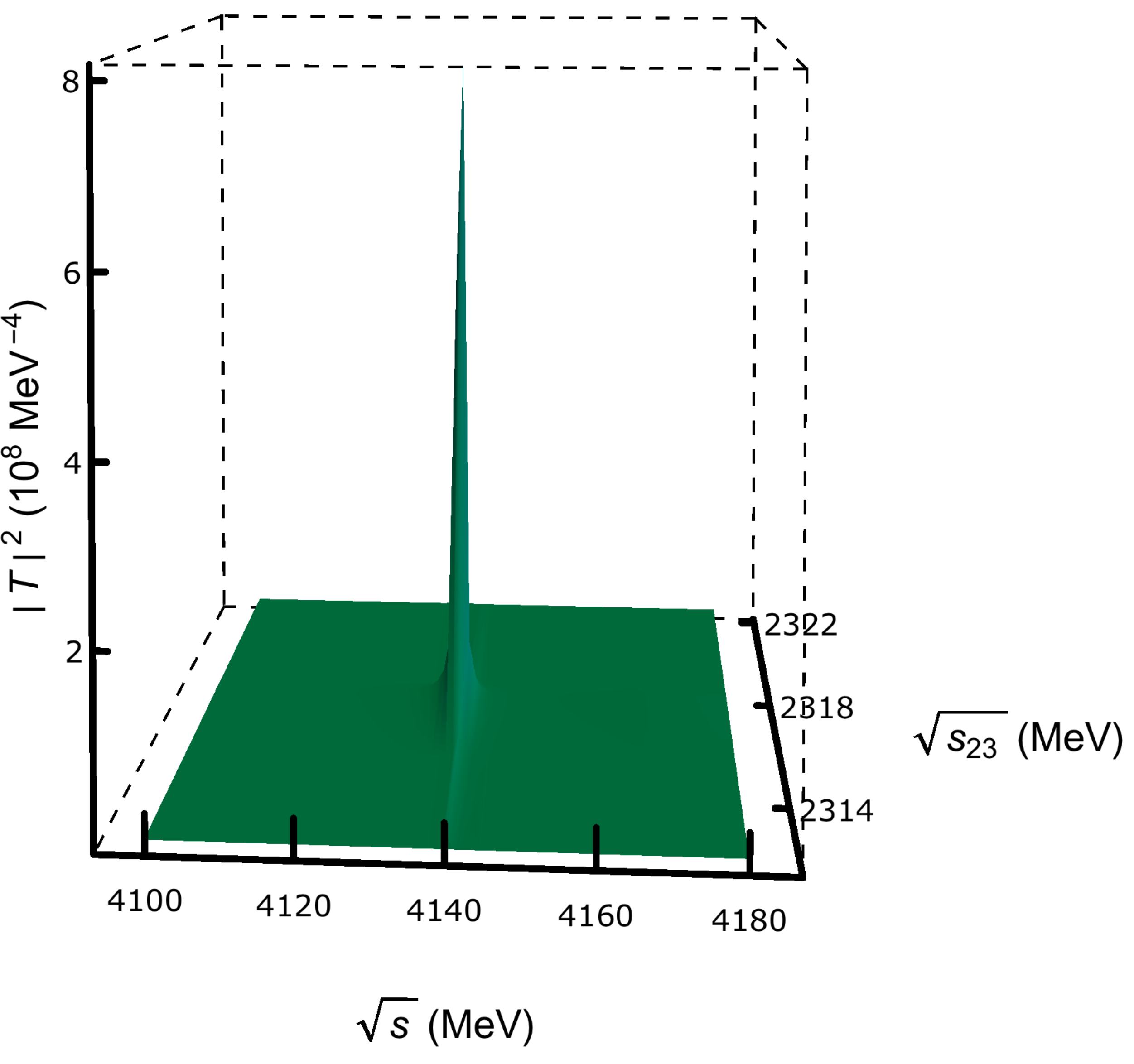}
\caption{Modulus squared of the $T_R$-matrix related to the process $DDK\to DDK$ in the $(I,I_{23})=(1/2,0)$ configuration.}\label{fig}
\end{figure}

In Fig.~\ref{fig} we show the modulus squared three-body amplitude, $|T_R|^2$, for the process $DDK\to DDK$ for total isospin $I=1/2$ and $I_{23}=0$, as a function of the energy of the three-body system, $\sqrt{s}$, and the invariant mass $\sqrt{s_{23}}$ of one of the $DK$ subsystems. 
As can be seen, a peak at $\sqrt{s}=4140$ MeV is found when the invariant mass of the $DK$ subsystem in isospin 0 is $\sim 2318$ MeV, which corresponds to the mass of the $D^*_{s0}(2317)$ formed in the subsystem. This result is in line with the one found in Ref.~\cite{SanchezSanchez:2017xtl}, in which the two body $D-D^*_{s0}(2317)$ system was studied without explicitly considering the three-body dynamics involved. Note that in Ref.~\cite{SanchezSanchez:2017xtl} two descriptions were taken into account for $D^*_{s0}(2317)$: as a compact $c\bar{s}$ state and as a $DK$ bound state. In both cases, predictions for the existence of a $D-D^*_{s0}$ state were made. However, as mentioned by the authors, the uncertainty involved in the former description is larger than in the latter case. In the present work, we have considered that the properties of $D^*_{s0}(2317)$ are predominantly understood in terms of the $DK$ and $D_s\eta$ interactions, as indicated from recent lattice studies and theoretical calculations~\cite{Lang:2014yfa, Torres:2014vna, Bali:2017pdv, Albaladejo:2018mhb}.

The result shown in Fig.~\ref{fig} implies that a state with charm 2, strangeness $+1$, and isospin $1/2$ is formed as a consequence of the dynamics involved in the system. It is interesting to notice that the $DD$ pair alone do not form a bound state, but adding a Kaon to the system binds it and produces an exotic meson with double charm, nonzero strangeness. If we denote the state found here as $R^{++}$, by the isospin symmetry, its charge +1 partner, $R^{+}$, with the third component of isospin $-1/2$, should also exist in nature.

It should be mentioned that even though the state found here is a bound state, the incorporation of two-body channels, open for decay, when coupled to the three-body channels considered, could lead to a width of few MeV. A small width is expected even if there is a phase space available for decay to two-body channels, like, $D^+_s D^{*+}$, $D^+ D^{*+}_s$, since this dynamics is not crucial for the generation of the three-body state found and, for this reason, its coupling to two-body channels should be very small.

Before discussing further properties of this state, we study how solid our findings are. For this, we investigate the sensitivity of the results to the parameters of the model, which are basically the subtraction constants used to regularize the loop functions when solving the Bethe-Salpeter equation to get the two-body $t$-matrices of the subsystems. While the subtraction constant used to regularize the $DK$, $D_s\pi$ and $D_s\eta$ loops has been fixed to reproduce the properties of the $D^*_{s0}(2317)$,  the situation is different for the $DD$ and $DD_s$ interactions. The dynamics involved in such double charm systems does not give rise to any state and the value of the subtraction constants is taken to be same as those used to reproduce data and properties of states found in other charm sectors (such as $X(3700)$, arising from the $D\bar D$ interaction~\cite{Zhang:2006ix,Gamermann:2006nm,Gamermann:2007bm}, or $X(3872)$, generated from the $D\bar D^*$ dynamics). In that sense, we could vary the subtraction constant used to regularize the $DD$ and $DD_s$ loops and check the dependence of the peak found in the three-body system. Varying this subtraction constant in a reasonable range, from $-1.3$ to $-1.5$, produces small changes in the magnitude of the three-body $T$-matrix, while the peak position remains basically unaltered. On the other hand, the formation of the $D^*_{s0}(2317)$ in the $DK$ subsystem and coupled channels is found to be essential to guarantee the formation of the three-body state. For example, a reduction in the strength of the interaction which leads to the generation of $D^*_{s0}(2317)$ by about 10\% is enough to weaken the attraction in the $DK$ subsystem and coupled channels and to break open the three-body state. 

Let us now discuss some properties, other than the quantum numbers, of the three-body bound state found in this work. One relevant property is the size of such an exotic state. It is important to know if the state is compact, since the interaction in two subsystems is attractive. Or, knowing that adding a charm meson to a Kaon produces $D^*_{s0}(2317)$, which is a molecule like state, does adding a $D$ to such a system leads to an extended object. One way to answer this question would be to solve the Faddeev equations in the configuration space, as done in Refs.~\cite{Hiyama:2003cu,Kanada-Enyo:2008wsu,Torres:2011jt}, which, however, is out of the scope of this work. Alternatively, we could treat the state found here as a $D-D^*_{s0}(2317)$ state of mass $M_R$ to estimate the mean square distance among the constituent hadrons. For this, following Refs.~\cite{Gamermann:2009uq,YamagataSekihara:2010pj,Aceti:2012dd}, on one hand, we can write the wave function $\left<\vec{x}|\psi\right>$ of the state generated as a consequence of the $DD^*_{s0}$ dynamics, as
\begin{align}
\left<\vec{x}|\psi\right>=\alpha\sqrt{\frac{2}{\pi}}\frac{1}{r}\,\text{Im}\left[\int\limits_0^\Lambda dp\, p\frac{e^{ipr}}{M_R-M_D-M_{D^*_{s0}}-\frac{p^2}{2\mu}}\right],\label{wave}
\end{align}
with $\mu$ being the reduced mass of the system. On the other hand, we can write the $DD^*_{s0}$ $T$-matrix in the Breit-Wigner form as
\begin{align}
T_{DDs}(s)=\frac{g^2}{s-M^2_R+i\Gamma_R M_R},\label{TBW}
\end{align}
with $\Gamma_R$ being the width of the state and here $s$ corresponds to the center of mass energy of the $DD^*_{s0}$ system. The quantum mechanical coupling $\alpha$ in Eq.~(\ref{wave}) and the field theoretical coupling $g$ in Eq.~(\ref{TBW}) are related through
\begin{align}
g^2&=-\left[\frac{dG}{ds}\Bigg|_{s=M^2_R}\right]^{-1}=64\pi^3\mu B^2\alpha^2\label{g2},
\end{align}
where $G(s)$ and $B$ are, respectively, the loop function and binding energy (with respect to the $D-D^*_{s0}(2317)$ threshold) of the $DD^*_{s0}(2317)$ system,
\begin{align}
G(s)&=\int\limits_0^\Lambda \frac{dp}{(2\pi)^2} p^2\frac{E_D+E_{D^*_{s0}}}{E_D E_{D^*_{s0}}\left[s-(E_D+E_{D^*_{s0}})^2+i\epsilon\right]},\label{G}\\
E_i&=\sqrt{p^2+M^2_i}.
\end{align}
In Eq.~(\ref{G}), $\Lambda\sim700-1000$ MeV corresponds to the cut-off used to regularize the $DD^*_{s0}(2317)$ loop $G(s)$ of Eq.~(\ref{G}) when solving the scattering problem $DD^*_{s0}(2317)\to D D^*_{s0}(2317)$. As discussed in Ref.~\cite{Gamermann:2009uq}, the value of $\alpha$ obtained from Eq.~(\ref{g2}) has a very smooth dependence on the cut-off $\Lambda$, so it is mostly determined by the binding energy.

Using the wave function in Eq.~(\ref{wave}), and varying $\Lambda\sim 700-1000$ MeV, we can determine the mean square distance $\left<r^2\right>$ for the system, and we get
\begin{align}
\sqrt{\left<r^2\right>}\sim 1.0-1.4~\text{fm}.\label{r2DK}
\end{align}
This result when compared with the mean square distance for the $DK$ bound state $D^*_{s0}(2317)$, $\sqrt{\left<r^2\right>}\sim 0.7~\text{fm}$~\cite{Cho:2011ew}, is about 1.4-2 times larger. We can also compare Eq.~(\ref{r2DK}) with the corresponding value obtained in Ref.~\cite{SanchezSanchez:2017xtl}, $\sim 1.0-1.6$ fm, and conclude that both results are compatible.

\begin{figure}[h!]
\centering
\includegraphics[width=0.3\textwidth]{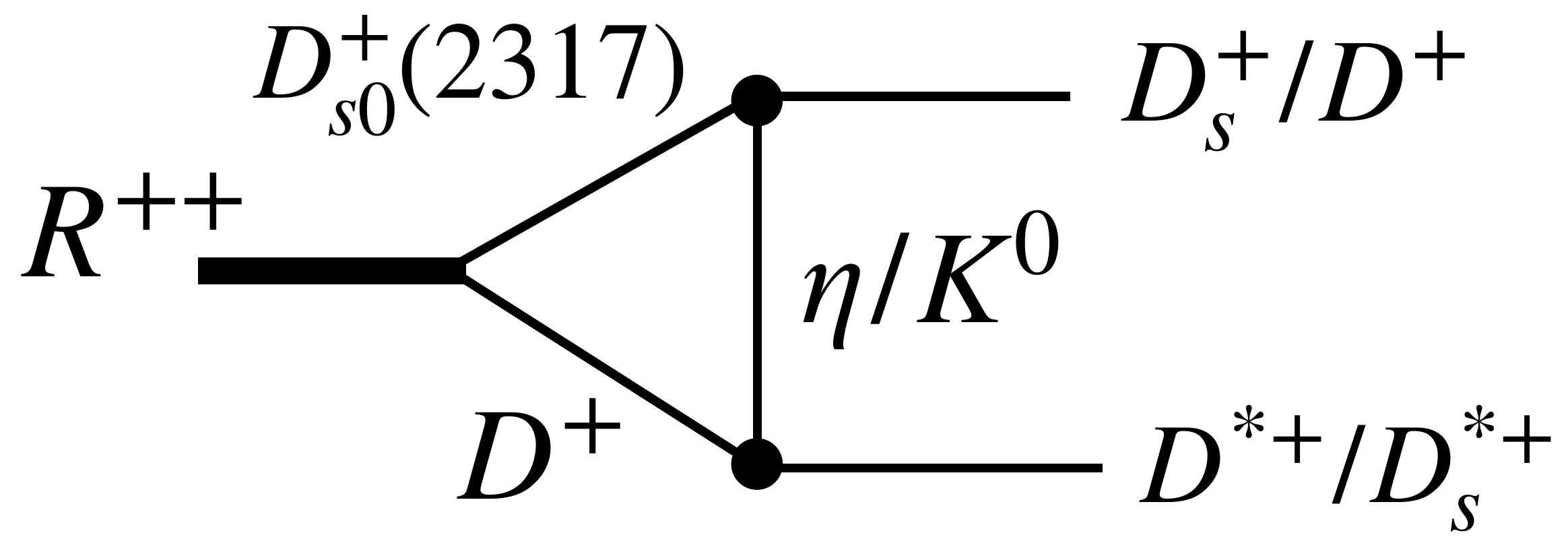}
\caption{A decay mechanism for the state found in this work.}\label{trianfig}
\end{figure}

A question might arise about the possibility of experimental investigations of the state found in the present work and how its three-body nature can be confirmed in experiments. The recent detection of a charm $+2$ baryon by the LHCb collaboration~\cite{Aaij:2017ueg}, and the search of the double charm tetraquark $T_{cc}$ state in heavy ion collisions~\cite{Cho:2017dcy}, indicate that the detection of the $R^{++}$ state can be accomplished. A signal for the state $R^{++}$ should be looked for in systems like $D^+_s D^{*+}$, $D^+ D^{*+}_s$, since it can decay to such channels, as shown in Fig.~\ref{trianfig}, or in three-body channels like $D D_s\gamma$, $D D_s\pi$, where the invariant mass of $D_s\gamma$ and $D_s\pi$ should be compatible with the formation of the $D^*_{s0}(2317)$. 
These decay widths depend on the underlying structure of the decaying particle through the coupling constants of $R^{++}$ to $D^*_{s0}(2317) D$ and of $D^*_{s0}(2317)$ to $DK$, $D_s\eta$, $D\gamma$, etc., thus, the values obtained for the widths will be a clear projection of the underlying three-body dynamics considered in the present work. A theoretical calculation of such processes is currently in progress and should be reported shortly. Similarly, the three-body nature of the $R^{++}$ state has its implications on the size of the state, and as we have shown in this work, the mean square distance can be around a factor $1.4-2$ bigger than that of $D^*_{s0}(2317)$. The size of $R^{++}$  can be investigated by determining the value of the production yield of the state in heavy ion collisions, where molecular states have bigger production yields as compared to compact bound quark states~\cite{Cho:2017dcy}.  A precise determination of the production yield of $R^{++}$ in heavy ion collisions should also be obtained in future works.

Finally, we must mention that we have also calculated the total isospin $3/2$ three-body $T$-matrix and we find no states formed.

\section{Conclusions}
We have studied the $DDK$ system and coupled channels by solving the Faddeev equations and calculating the three-body scattering matrix. We have found that an isospin 1/2 state is formed at 4140 MeV when the $D^*_{s0}(2317)$ is generated in the $DK$ subsystems. Such a result is compatible with the one found in Ref.~\cite{SanchezSanchez:2017xtl} where the system $D-D^*_{s0}(2317)$ was studied without considering explicit three-body dynamics. A state with these quantum numbers (spin-parity $0^-$, isospin 1/2, charm $+2$, strangeness $+1$) has not been observed so far. We hope that our results motivate its search in experimental investigations.

\section{Acknowledgment}
This work was partly supported the National Natural Science Foundation of China (NSFC) under Grants Nos. 11735003 and 11522539, and Conselho Nacional de Desenvolvimento Cient\'ifico e Tecnolo\'ogico (CNPq) under Grant Nos. 310759/2016-1 and 311524/2016-8. A.M.T and K.P.K thank Beihang University for the hospitality during their stay when this work was initiated.

\appendix
\section{Coefficients for the $DD$ amplitudes}\label{coef}
In this appendix we give the coefficients appearing in Eq.~(\ref{Cs}).
\begin{table}[h!]
\caption{$A_k$ coefficients for different processes. The exchanged vector meson is indicated next to the coefficient.}
\vspace{0.2cm}
\begin{tabular}{c|cc}
&$D^+D^0$&$D^0D^+$\\
\hline\\
$D^+D^0$& $\begin{array}{c}1/2~(\omega)\\ -1/2~(\rho^0)\\1~(J/\psi)\end{array}$&$1~(\rho^+)$ \\
$D^0D^+$& $1~(\rho^-)$&$\begin{array}{c}1/2~(\omega)\\ -1/2~(\rho^0)\\1~(J/\psi)\end{array}$
\end{tabular}
\quad
\vspace{0.5cm}
\begin{tabular}{c|c}
&$D^+D^+$\\
\hline\\
$D^+D^+$& $\begin{array}{c}1/2~(\omega)\\ 1/2~(\rho^0)\\1~(J/\psi)\end{array}$
\end{tabular}
\begin{tabular}{c|cc}
&$D^+_sD^0$&$D^0D^+_s$\\
\hline\\
$D^+_sD^0$& $1~(J/\psi)$&$1~(K^{*+})$ \\
$D^0D^+_s$& $1~(K^{*-})$&$1~(J/\psi)$
\end{tabular}
\quad
\begin{tabular}{c|cc}
&$D^+_sD^+$&$D^+D^+_s$\\
\hline\\
$D^+_sD^+$& $1~(J/\psi)$&$1~(K^{*0})$ \\
$D^+D^+_s$& $1~(\bar K^{*0})$&$1~(J/\psi)$
\end{tabular}
\end{table}
\begin{table}
\caption{$B_k$ coefficients for different reactions. We indicate the exchanged vector meson next to the coefficient.}
\vspace{0.2cm}
\begin{tabular}{c|cc}
&$D^+D^0$&$D^0D^+$\\
\hline\\
$D^+D^0$& $1~(\rho^+)$& $\begin{array}{c}1/2~(\omega)\\ -1/2~(\rho^0)\\1~(J/\psi)\end{array}$ \\
$D^0D^+$& $\begin{array}{c}1/2~(\omega)\\ -1/2~(\rho^0)\\1~(J/\psi)\end{array}$&$1~(\rho^-)$\\
\end{tabular}
\quad
\vspace{0.5cm}
\begin{tabular}{c|c}
&$D^+D^+$\\
\hline\\
$D^+D^+$& $\begin{array}{c}1/2~(\omega)\\ 1/2~(\rho^0)\\1~(J/\psi)\end{array}$
\end{tabular}
\begin{tabular}{c|cc}
&$D^+_sD^0$&$D^0D^+_s$\\
\hline\\
$D^+_sD^0$& $1~(K^{*+})$&$1~(J/\psi)$ \\
$D^0D^+_s$& $1~(J/\psi)$&$1~(K^{*-})$
\end{tabular}
\quad
\begin{tabular}{c|cc}
&$D^+_sD^+$&$D^+D^+_s$\\
\hline\\
$D^+_sD^+$& $1~(K^{*0})$&$1~(J/\psi)$ \\
$D^+D^+_s$& $1~(J/\psi)$&$1~(\bar K^{*0})$
\end{tabular}
\end{table}
\clearpage
\section{Off-shell contribution and three-body contact terms}\label{cancell}
\begin{figure}[h]
\begin{center}
\includegraphics[width=0.9\textwidth]{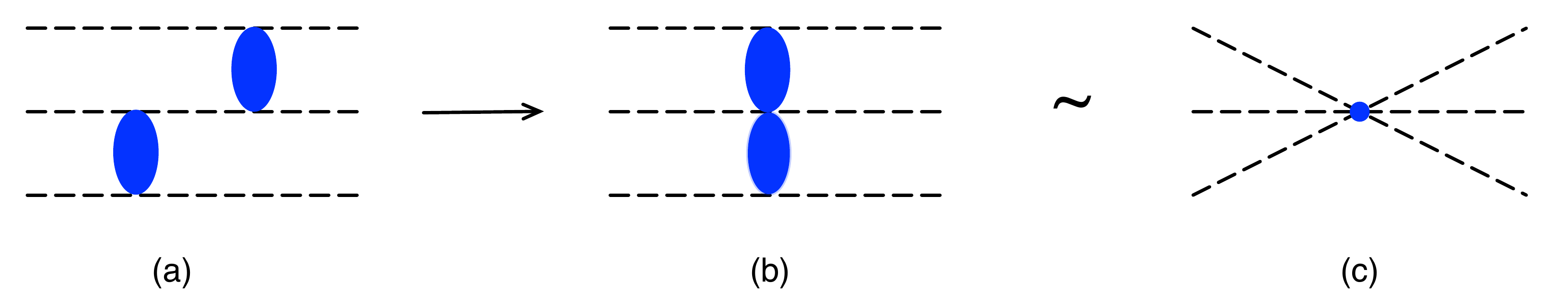}
\caption{(a) A lowest order diagram contributing to Faddeev equations. (b) The same diagram, when considering the off-shell parts of the two-body amplitudes, which cancel a propagator in the three-body diagram and leads to a three-body contact interaction shown in (c).}\label{contactdiag}
\end{center}
\end{figure}

In this appendix we investigate if diagrams other than the kind shown in Fig.~\ref{T1fig} can contribute to the three-body interactions being studied here, such as three-body contact interactions.  As was shown in Refs.~\cite{MartinezTorres:2007sr,Khemchandani:2008rk,MartinezTorres:2008gy,Torres:2011jt}, three-body contact interactions can arise from the Lagrangians used to determine the two-body amplitudes needed to solve the Faddeev equations. It was further discussed that there are yet other sources of such contact terms, which arise from the off-shell parts of the input two-body amplitudes, which analytically cancel the propagator in the three-body diagrams at the tree level, leading to diagrams which are topologically equivalent to a three-body  contact term (see Fig.~\ref{contactdiag}).  
An exact analytical cancellation was found in the SU(3) limit in Refs.~\cite{MartinezTorres:2007sr, MartinezTorres:2008gy} for systems made of two pseudoscalar mesons and an octet baryon or of two pseudoscalars and a vector meson. While in the case of a study of three pseudoscalars, such a cancellation was found in the chiral limit~\cite{Torres:2011jt}. The cancellation among such contributions, in realistic cases, is not exactly null but the sum of all of them has been found to be negligible, when compared with the results obtained by calculations done by considering the on-shell  parts of the two-body $t$-matrices. Thus, it has been shown that one can neglect such sources of three-body contact interactions and work just with the on-shell part of the two-body $t$-matrices.

In the present case, we study a system made of two heavy and a light meson and the scattering amplitudes for the two-body subsystems have been obtained by solving Bethe-Salpeter equations with the kernels derived from Lagrangians based on the chiral and the heavy quark symmetry.  Such a Lagrangian does not provide us with a three-body contact term at the same order. Nevertheless, as mentioned above, contact terms can arise from other sources and we can analise them. It is also possible to make such an analysis by obtaining amplitudes from the SU(4) Lagrangian, which does provide a contact interaction term with six fields. Lagrangians based on both symmetries, SU(4) and the heavy quark, have been used to study different meson-meson interactions and similar results have been found. For instance, the properties of $D_s(2317)$ have been successfully described in Refs.~\cite{Gamermann:2006nm} by working with the SU(4) symmetry and in
Refs.~\cite{Guo:2006fu,Guo:2017jvc} by using the heavy quark symmetry. In the present work we present an analysis of three-body contact interactions coming from both type of Lagrangians.

\subsection{Interactions taken from the SU(4) Lagrangian}
\begin{figure}[ht!]
\begin{center}
\includegraphics[width=0.7\textwidth]{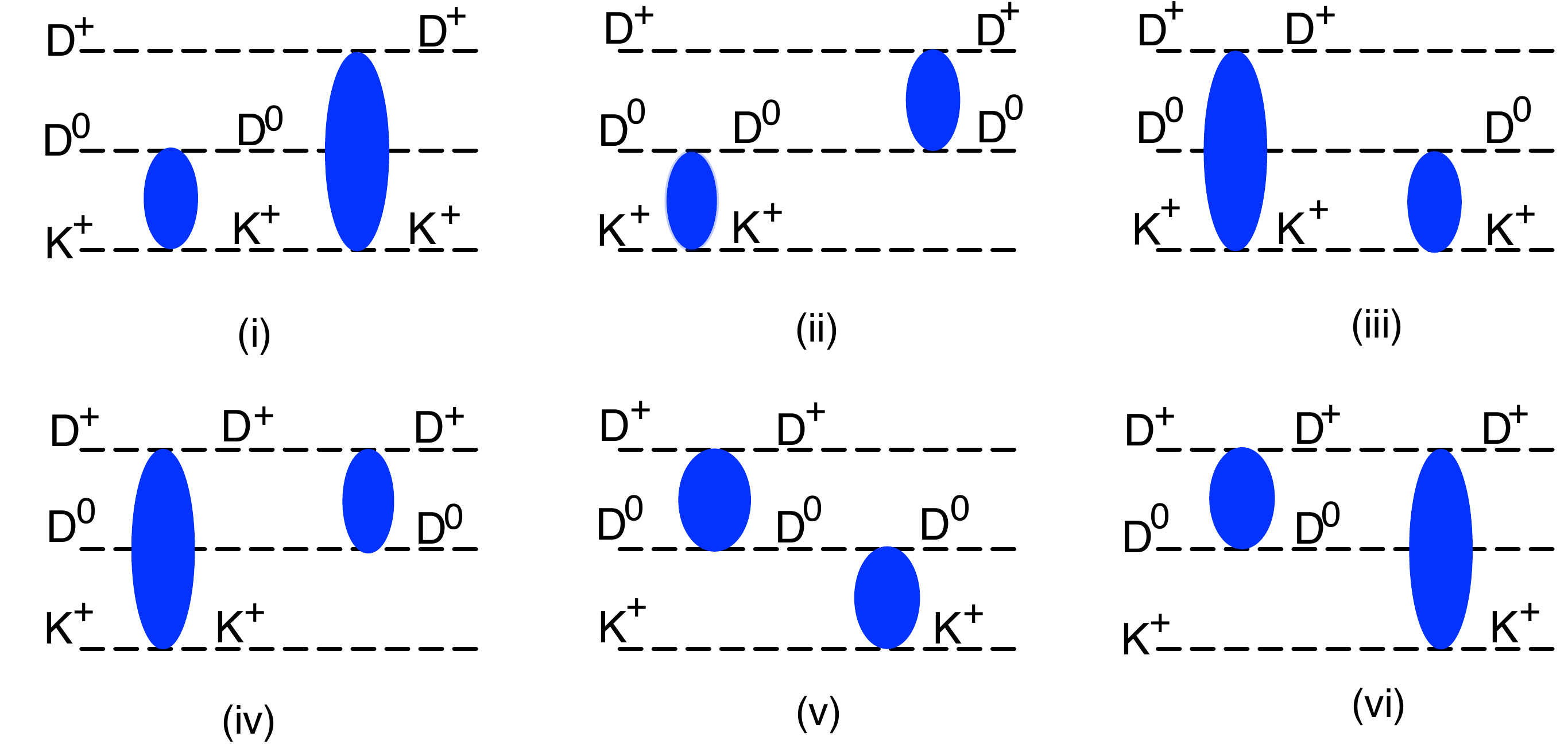}
\caption{Three-body interaction diagrams contributing to the Faddeev equations at the lowest order.}\label{contactdiag2}
\end{center}
\end{figure}

We take the general Lagrangian from Ref.~\cite{Torres:2011jt} 
\begin{equation}
\mathcal{L} = \frac{f^2}{4} Tr\left[ \partial_\mu U^\dagger \partial^\mu U + M(U + U^\dagger) \right],
\end{equation}
where 
\begin{equation}
U = e^{i\sqrt{2} \phi/f},
\end{equation}
with the $\phi$-matrix being as defined by Eq.~(\ref{phi2})  of the manuscript and 
\begin{equation}
M = \left(\begin{array}{cccc}
m_\pi^2&0&0&0\\
0&m_\pi^2&0&0\\
0&0&2m_K^2-m_\pi^2&\\
0&0&0&2 m_D^2 - m_\pi^2
\end{array}
\right).
\end{equation}
The $3 \phi \to 3 \phi$ contact term is obtained as
\begin{equation}
V_{3\phi \to 3\phi} = -\frac{1}{180 f^4} \langle 6 \partial^\mu \phi \phi^2 \partial_\mu\phi \phi^2 + 2 \partial^\mu \phi \phi^4 \partial_\mu\phi - 8 \partial^\mu \phi \phi^3 \partial_\mu\phi \phi - M\phi^6\rangle.\label{3bodyV}
\end{equation}
To analise different three-body contact interactions, let us consider the $D^+ D^0 K^+$ channel, as an example. Assigning four momenta to particles in the process as: $D^+(K_1) D^0 (K_2) K^+ (K_3) \to D^+ (K_1^\prime) D^0 (K_2^\prime) K^+ (K_3^\prime)$, the amplitude is obtained from Eq.~(\ref{3bodyV}) as 
\begin{eqnarray}
V^{1, \,\text{contact}}_{D^+D^0 K^+}&=-\dfrac{1}{180 f^4}\biggl[12\left( K_2 K_2^\prime -K_1K_3-K_1^\prime K_3^\prime \right) + 2 \left(- K_2^\prime K_3^\prime - K_2 K_3 + K_1K_1^\prime \right.\biggr. \nonumber\\
&+ \left. K_3 K_3^\prime +K_1^\prime K_2 + K_2^\prime K_1 \right) - 8\left(K_2^\prime K_3 + K_1^\prime K_3 - K_1^\prime K_2^\prime +K_2 K_3^\prime\right.\nonumber\\
&-\left. K_1K_2 + K_1 K_3^\prime \right) -2\left(2 m_D^2 + m_K^2\right) \biggl. \biggr],
\end{eqnarray}
where the superscript indicates the first kind of contact term considered. We shall now discuss the three-body contact terms arising from the situation described in Fig.~\ref{contactdiag}. The diagrams contributing to the three-body interactions, at the lowest order, are  shown in Fig.~\ref{contactdiag2}, for $D^+D^0 K^+ \to D^+D^0 K^+$.
To evaluate these diagrams we need the following amplitudes (taken from Ref.~\cite{Gamermann:2006nm})
\begin{eqnarray}
V_{D^+ D^0 \to D^+ D^0} &=& -\frac{1}{6 f^2}\biggl( -\frac{3}{2} \left(1 + \psi_5\right) s_{DD} + 2 \left(2 + \psi_5\right) m_D^2  \biggr.\nonumber\\
&+& \left(\frac{1-\psi_5}{2}\right) \left(t_{DD} -u_{DD}\right)+ \frac{1+\psi_5}{2}\sum\limits_{i}\left(P_i^2 - m_i^2\right)\biggl.\biggr), \label{VDD}\\
V_{D^+ K^0 \to D^+ K^0} &=& -\frac{1}{6 f^2}\biggl( -\frac{3}{2} s_{DK} + \left(\gamma +\frac{1}{2}\right) \left(t_{DK} - u_{DK} \right) \biggr.\nonumber\\ 
&-& \biggl. \frac{1}{2}\sum\limits_{i}\left(K^2_i - m_i^2\right)\biggr),\label{VDK}\\
V_{D^+ K^+ \to D^+ K^+} &=& 0, 
\end{eqnarray}
where $s_{ij}$, $t_{ij}$, $u_{ij}$ represent the Mandelstam variables for the $ij$ system, $P_i$  ($K_i$) is the four momentum of the $i$th particle in the $D^+ D^0 \to D^+ D^0$  ($D^+ K^0 \to D^+ K^0$) process and $\psi_5$, $\gamma$ are defined, in Ref.~\cite{Gamermann:2006nm}, as
\begin{eqnarray}
\psi_5 = - \frac{1}{3} + \frac{4}{3}\left(\frac{m_L}{m_{J/\Psi}}\right)^2,~
\gamma = \left(\frac{m_L}{m_H}\right)^2,
\end{eqnarray}
with $m_L$ ($m_H$) being the mass of  light (heavy) vector meson. These variables were introduced in Ref.~\cite{Gamermann:2006nm} to incorporate the breaking of the SU(4) symmetry for the scattering of two mesons. However, to be consistent with the contact term obtained from Eq.~(\ref{3bodyV}), we will eventually use the SU(4) limit for their values.

\begin{figure}[ht!]
\begin{center}
\includegraphics[width=0.5\textwidth]{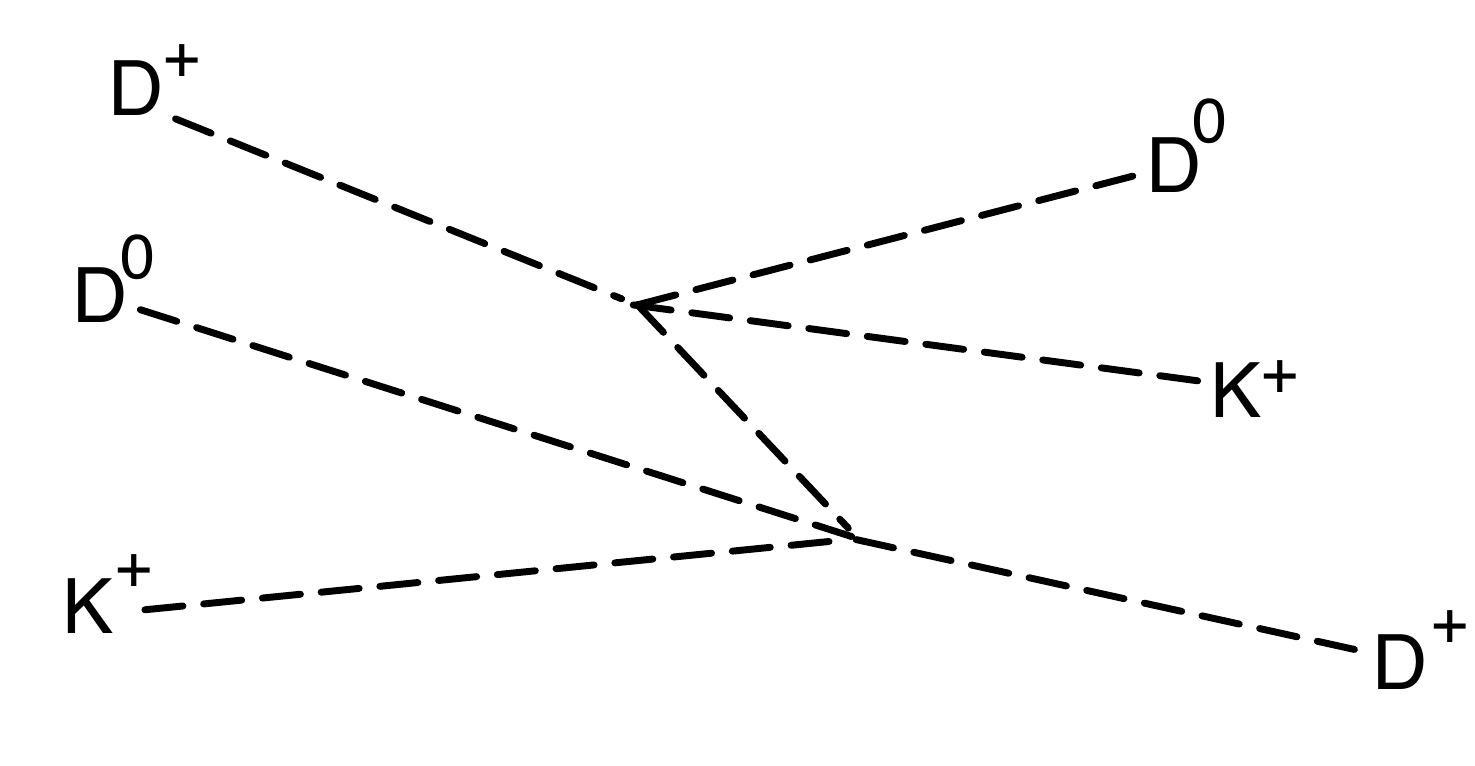}
\caption{Another source of a three-body contact interaction.}\label{contactdiag3}
\end{center}
\end{figure}

It can be seen that the off-shell parts of the  two-body amplitudes in Eqs.~(\ref{VDD}), (\ref{VDK}) go as $(q_i^2 - m_i^2)$, which precisely cancel a propagator in the three-body diagrams shown in Fig.~\ref{contactdiag2}, and give rise to a three-body contact term.
Using the amplitudes in Eqs.~(\ref{VDD}), (\ref{VDK}), we obtain such contact  terms as: 
\begin{eqnarray}
&&V^{2{\textrm i}, \,\text{contact}} = V^{2\textrm{iii}, \,\text{contact}}= V^{2\textrm{iv}, \,\text{contact}}= V^{2\textrm{vi}, \,\text{contact}} =0,\nonumber \\\nonumber\\
&&V^{2\textrm{ii}, \,\text{contact}}_{D^+D^0 K^+} = \frac{1}{72 f^4} \Biggl\{ \left(\frac{1+\psi_5}{2}\right)\left[3 \left(K_2 + K_3\right)^2 +  2\left( \gamma + \frac{1}{2}\right)\right. \Biggr. \nonumber\\
&\times& \biggl. \biggl((K_3- K_3^\prime)^2 - (K_2 - K_3^\prime)^2\biggr) \biggr] -\left[- \frac{3}{2}\left(1+ \psi_5 \right) (K_1^\prime+K_2^\prime)^2\right. \nonumber\\
&+& 2(2+\psi_5) m_D^2 + \left(\frac{1-\psi_5}{2} \right)\biggl((K_1- K_1^\prime)^2 - (K_1 - K_2^\prime)^2 \biggr) \nonumber\\
&+& \left(\frac{1+\psi_5}{2} \right) \bigl((K_1^\prime+K_2^\prime-K_1)^2-m_D^2\bigr) \biggl.\biggr]\Biggl. \Biggr\},
\end{eqnarray}
\begin{eqnarray}
&&V^{2{\textrm v}, \,\text{contact}}_{D^+D^0 K^+} = \frac{1}{72 f^4} \Biggl\{ \left(\frac{1+\psi_5}{2}\right)\left[3 \left(K_2^\prime + K_3^\prime \right)^2 +  2\left( \gamma + \frac{1}{2}\right)\right. \Biggr. \nonumber\\
&\times& \biggl. \biggl((K_3- K_3^\prime)^2 - (K_3 - K_2^\prime)^2\biggr) - (K_2^\prime+K_3^\prime-K_3)^2 + m_D^2 \biggr] \nonumber\\
&-&\left[ -\frac{3}{2}\left(1+ \psi_5 \right) (K_1+K_2)^2 + 2(2+\psi_5) m_D^2 \right.\nonumber\\
&+& \left(\frac{1-\psi_5}{2} \right)\biggl((K_1- K_1^\prime)^2 - (K_2 - K_1^\prime)^2 \biggr)  \biggl.\biggr]\Biggl. \Biggr\}.
\end{eqnarray}
Before proceeding further,  as noticed in Ref.~\cite{Torres:2011jt}, we recall that more diagrams may exist which can contribute to three-body contact interactions.  An additional diagram exists in the present case, as shown in Fig.~\ref{contactdiag3}.
In this case too, the off-shell parts of the input two body interactions possess the feature which kills a  propagator in the three-body diagram, leading to yet another three-body contact term. Since the purpose is to sum all kinds of three-body contact interactions, we take into account this latter term too. The on-shell contribution of the two-body amplitude to this diagram is much smaller than those shown in Fig.~\ref{contactdiag2}, since a larger momentum transfer is involved in this case, and we neglect it.  
We obtain the contribution of the off-shell part of the two-body amplitudes to the diagram in Fig.~\ref{contactdiag3} as
\begin{eqnarray}
&&V^{3, \,\text{contact}}_{D^+D^0 K^+} =- \frac{1}{72 f^4} \Biggl\{\frac{3}{2} \left( K_2^\prime + K_3^\prime \right)^2 + \left( \gamma +\frac{1}{2} \right) \left[(K_1- K_2^\prime)^2 -(K_1 - K_3^\prime)^2 \right] \Biggr. \nonumber\\
&+& \frac{3}{2} \left( K_2 + K_3 \right)^2 + \left( \gamma +\frac{1}{2} \right) \left[(K_2- K_1^\prime)^2 -(K_3 - K_1^\prime)^2\right] - \frac{t - m_K^2}{2}  \Biggl.  \Biggr\},
\end{eqnarray}
where $t = (K_2+K_3 - K_1^\prime)^2$.

For the sake of an analytical evaluation of all these amplitudes, we consider the three-momenta of the external particles to be negligible as compared to their energies for a total energy near the resonance mass, and we use
\begin{eqnarray}
K_1^0 &=& \frac{s - s_{23} +m_1^2}{2 \sqrt{s}},\nonumber\\
K_2^0 &=& \frac{s - s_{13} +m_2^2}{2 \sqrt{s}},\nonumber\\
K_3^0 &=& \frac{s - s_{12} +m_3^2}{2 \sqrt{s}}.\label{approx1}
\end{eqnarray}
Further, as found in the present work, around the three-body resonance, the invariant masses of both $DK$ pairs have a value of $M_{2R}\simeq M_{D_s(2317)}$. Thus, we can write
\begin{eqnarray}
s_{23} \simeq s_{13} = M^2_{2R} \simeq (m_D + m_K - \Delta E)^2,\label{approx2}
\end{eqnarray}
where $\Delta E$ is the modulus of the $DK$ binding energy $\sim$ 50 MeV, and
\begin{eqnarray}
s_{12} = s + 2 m_D^2 + m_K^2 -  s_{13} -  s_{23},\label{approx3}
\end{eqnarray}
with $s$ representing the Mandelstam variable for the $DDK$ system. We take $\sqrt{s}$ to be the mass of the resonance, $M_{3R} = M_{2R} + m_K - \Delta E^\prime$, where $\Delta E^\prime \sim 50$ MeV is the binding energy of the $DD^*_{s0}$ system. Further, as mentioned earlier, we consider the SU(4) limit 
\begin{equation}
\psi \to 1,~\gamma \to 1.\label{approx4}
\end{equation}
As a result, we obtain
\begin{eqnarray}
&&V^{1, \,\text{contact}}_{D^+D^0 K^+}+V^{2\textrm{ii}, \,\text{contact}}_{D^+D^0 K^+}+V^{2\textrm{v}, \,\text{contact}}_{D^+D^0 K^+}+V^{3, \,\text{contact}}_{D^+D^0 K^+} \nonumber\\
&=&   \frac{m_D m_K}{2 f^4}\Biggr( 1- \frac{(37 m_D^2 +  37 m_K m_D + 154 m_K^2)\Delta E^\prime}{90 \left(2 m_D + m_K\right) m_D m_K} + O\left(\frac{\Delta E^{\prime 2}}{m_Dm_K}\right)\Biggl)\nonumber\\
&+&\frac{ m_Dm_K \Delta E}{90f^4} \Biggr(\frac{32 m_K - 45 m_D}{m_D m_K} +\frac{(37 m_D^2 + 616 m_K m_D + 154 m_K^2)\Delta E^\prime}{2m_D m_K \left(2 m_D + m_K\right)^2} + O\left(\frac{\Delta E^{\prime 2}}{m_D m_K}\right)\Biggl)\nonumber\\
&+&O\left(\frac{\Delta E^2}{m_D m_K}\right),\label{3b1}
\end{eqnarray}
which is, approximately, 
\begin{equation}
V^{1, \,\text{contact}}_{D^+D^0 K^+}+V^{2\textrm{ii}, \,\text{contact}}_{D^+D^0 K^+}+V^{2\textrm{v}, \,\text{contact}}_{D^+D^0 K^+}+V^{3\textrm{i}, \,\text{contact}}_{D^+D^0 K^+}+V^{3\textrm{ii}, \,\text{contact}}_{D^+D^0 K^+} \sim \frac{m_D m_K}{2 f^4}. \label{app}
\end{equation}
This contribution becomes null in the limit of massless light quarks $m_u, m_d, m_s = 0$, which is the chiral limit. This finding is  similar to the one found in Ref.~\cite{Torres:2011jt} for a system of three light pseudoscalars. 

For a more realistic case, we can compare the result in Eq.~(\ref{3b1}) with the on-shell contribution of the two-body $t$-matrices to the diagrams shown in Fig.~\ref{contactdiag2}, which give the three-body amplitudes as
\begin{eqnarray}
&&V^{2\textrm{i}, \text{on}}_{D^+D^0 K^+} = V^{2\textrm{iii}, \text{on}}_{D^+D^0 K^+} = V^{2\textrm{iv}, \text{on}}_{D^+D^0 K^+} = V^{2\textrm{vi}, \text{on}}_{D^+D^0 K^+} = 0,\nonumber\\\nonumber\\
&&V^{2\textrm{ii}, \text{on}}_{D^+D^0 K^+} =\frac{1}{36 f^4}\frac{1}{(K_2+K_3 -K_3^\prime)^2-m_D^2}\Biggl\{-\frac{9(1+\psi_5)}{4}\left(K_1^\prime+K_2^\prime\right)^2 \Biggr.\nonumber\\
&\times& (K_2+K_3)^2 - \frac{3(1+\psi_5)}{2} \left(\gamma+\frac{1}{2}\right) \left(K_1^\prime+K_2^\prime\right)^2 \biggl[\left(K_3 - K_3^\prime\right)^2 -\left(K_2 - K_3^\prime\right)^2 \biggr]  \nonumber\\
&+& 3(2+\psi_5) m_D^2 (K_2+K_3)^2+ 2(2+\psi_5)\left(\gamma+\frac{1}{2}\right) m_D^2 \biggl[\left(K_3 - K_3^\prime\right)^2 - \left(K_2 - K_3^\prime\right)^2\biggr] \nonumber\\
&+&  \frac{3}{2}\left(\frac{1-\psi_5}{2}\right) (K_2+K_3)^2\biggl[\left(K_1 - K_1^\prime\right)^2 -\left(K_1 - K_2^\prime\right)^2 \biggr] + \left(\frac{1-\psi_5}{2}\right)\left(\gamma+\frac{1}{2}\right) \nonumber\\
&\times& \biggl[\left(K_1 - K_1^\prime\right)^2 -\left(K_1 - K_2^\prime\right)^2 \biggr] \biggl[\left(K_3 - K_3^\prime\right)^2 - \left(K_2 - K_3^\prime\right)^2\biggr] \Biggl. \Biggr\},
\end{eqnarray}
and
\begin{eqnarray}
&&V^{2\textrm{v}, \text{on}}_{D^+D^0 K^+} =\frac{1}{36 f^4}\frac{1}{(K_2^\prime+K_3^\prime -K_3)^2-m_D^2}\Biggl\{-\frac{9(1+\psi_5)}{4}\left(K_1+K_2\right)^2 \Biggr.\nonumber\\
&\times& (K_2^\prime+K_3^\prime)^2 - \frac{3(1+\psi_5)}{2} \left(\gamma+\frac{1}{2}\right) \left(K_1+K_2\right)^2 \biggl[\left(K_3 - K_3^\prime\right)^2 -\left(K_3 - K_2^\prime\right)^2 \biggr]  \nonumber\\
&+& 3(2+\psi_5) m_D^2 (K_2^\prime+K_3^\prime)^2+ 2(2+\psi_5)\left(\gamma+\frac{1}{2}\right) m_D^2 \biggl[\left(K_3 - K_3^\prime\right)^2 - \left(K_3 - K_2^\prime\right)^2\biggr] \nonumber\\
&+&  \frac{3}{2}\left(\frac{1-\psi_5}{2}\right) (K_2^\prime+K_3^\prime)^2\biggl[\left(K_1 - K_1^\prime\right)^2 -\left(K_2 - K_1^\prime\right)^2 \biggr] + \left(\frac{1-\psi_5}{2}\right)\left(\gamma+\frac{1}{2}\right) \nonumber\\
&\times& \biggl[\left(K_1 - K_1^\prime\right)^2 -\left(K_2 - K_1^\prime\right)^2 \biggr] \biggl[\left(K_3 - K_3^\prime\right)^2 - \left(K_3 - K_2^\prime\right)^2\biggr] \Biggl. \Biggr\}.
\end{eqnarray}
To make a consistent comparison, we use Eqs.~(\ref{approx1})$-$(\ref{approx4}) and deduce the above amplitudes up to the first order in $\Delta E$, $\Delta E^\prime$. In this way, we get
\begin{eqnarray}
&&V^{2\textrm{ii}, \text{on}}_{D^+D^0 K^+}+V^{2\textrm{v}, \text{on}}_{D^+D^0 K^+} = \frac{(2 m_D + m_K)}{4 m_K f^4 (- m_K \Delta E^\prime + m_K \Delta E + 2  m_D \Delta E)}\nonumber\\
&\times& \Biggl( 2\Delta E m_D^3 - 2\Delta E m_D^2 m_K + \frac{4 \Delta E^\prime m_D^3 m_K}{2m_D + m_K}- 2 m_D^3 m_K+ \frac{8 \Delta E^\prime m_D^2 m_K^2}{2 m_D + m_K}\Biggr)~~~~~~~~\label{vontotal}
\end{eqnarray}
 Using $\Delta E \sim \Delta E^\prime = 50$ MeV, and an average mass for kaon of 496 MeV, Eqs.~(\ref{3b1}) and (\ref{vontotal}) give, at the three-body resonance mass, \begin{eqnarray}
&&V^{1, \,\text{contact}}_{D^+D^0 K^+}+V^{2\textrm{ii}, \,\text{contact}}_{D^+D^0 K^+}+V^{2\textrm{v}, \,\text{contact}}_{D^+D^0 K^+}+V^{3, \,\text{contact}}_{D^+D^0 K^+}  \sim \frac{0.12 m_D^2}{f^4}~~~~~~~~~~~ \label{su4valoff}\\
&&V^{2\textrm{ii}, \text{on}}_{D^+D^0 K^+}+V^{2\textrm{v}, \text{on}}_{D^+D^0 K^+} \sim - \frac{19 m_D^2}{f^4} \label{su4valon}
\end{eqnarray}
It can be seen that the total amplitude coming from the three-body contact interactions is about 0.6$\%$ of  the amplitudes coming from the three-body interaction diagrams calculated with the on-shell two-body amplitudes, and it can, thus, be neglected. 

\subsection{Interactions based on the chiral and heavy quark symmetry}
In this case the contact interactions arise due to the cancellation of a propagator in the diagrams shown in Figs.~\ref{contactdiag2} and \ref{contactdiag3}, as a consequence of the structure of the off-shell part of two-body amplitudes. The relevant two-body interactions, here, obtained from Eq.~(\ref{heavy}), are
\begin{align}
\tilde V_{D^+ D^0 \to D^+ D^0} &= \frac{1}{8 f^2}\biggl( 3 s_{DD} - \left(t_{DD} - u_{DD} \right) - 4 m_D^2  + \sum\limits_{i}\left(P_i^2 - m_i^2\right)\biggr), \label{VDD2}\\
\tilde V_{D^+ K^0 \to D^+ K^0} &= -\frac{1}{8 f^2}\biggl( 3 s_{DK} +  \left(t_{DK} - u_{DK} \right) - 2 m_D^2 - 2 m_K^2\biggr.\nonumber\\
&\biggl.- \sum\limits_{i}\left(K^2_i - m_i^2\right)\biggr),\label{VDK2}\\
\tilde{V}_{D^+ K^+ \to D^+ K^+} &= 0. 
\end{align}
The contribution of the off-shell parts of these amplitudes to the diagrams in Figs.~\ref{contactdiag2} and \ref{contactdiag3} leads to the following amplitudes:
{\small
\begin{align}
\tilde V^{2{\textrm i}, \,\text{contact}}_{D^+D^0 K^+} &= \tilde V^{2\textrm{iii}, \,\text{contact}}_{D^+D^0 K^+} = \tilde V^{2\textrm{iv}, \,\text{contact}}_{D^+D^0 K^+}= \tilde V^{2\textrm{vi}, \,\text{contact}}_{D^+D^0 K^+} =0,\nonumber \\\nonumber\\
\tilde V^{2\textrm{ii}, \,\text{contact}}_{D^+D^0 K^+} &= \frac{1}{64 f^4} \Biggl\{ 2 m_D^2 + 6 K_1 K_2 + 2 K_1^\prime K_1-  2 K_1^\prime K_2 - 4 K_2^\prime K_3^\prime- 4 K_2^\prime K_3\Biggr\},\label{H1}\\
\tilde V^{2\textrm{v}, \,\text{contact}}_{D^+D^0 K^+}& = \frac{1}{64 f^4} \Biggl\{ 4 m_D^2 - 2 m_K^2 - 6 K_2 K_3 + 2 K_3^\prime (K_3-  K_2) + 8 K_1^\prime K_2^\prime-4 K_2^\prime K_1\Biggr\}, \label{H2}\\
\tilde V^{3, \,\text{contact}}_{D^+D^0 K^+}& = - \frac{1}{64 f^4}\Biggl\{ 2 m_D^2 + 6 K_2^\prime K_3^\prime - 2 K_1 K_2^\prime + 2 K_1 K_3^\prime + 4 K_2 K_3 + 4 K_1^\prime K_3 \Biggr\}.\label{H3}
\end{align}}
As in the previous subsection, we use Eqs.~(\ref{approx1})$-$(\ref{approx4}) and deduce the sum of the amplitudes in Eqs.~(\ref{H1})-(\ref{H3}) up to the first order in $\Delta E$, $\Delta E^\prime$, to get:
\begin{eqnarray}
&&\tilde V^{2\textrm{ii}, \,\text{contact}}_{D^+D^0 K^+}+\tilde V^{2\textrm{v}, \,\text{contact}}_{D^+D^0 K^+}+\tilde V^{3, \,\text{contact}}_{D^+D^0 K^+} =\frac{m_D(m_D - 2m_K)}{4 f^4} \nonumber\\
&+&\Delta E~\frac{(8 m_D - m_K)}{16 f^4} -3 \Delta E^\prime~\frac{(2 m_D^2 + 2 m_D m_K - 3 m_K^2)}{16 f^4(2 m_D +m_K)}, \label{Hoff}
\end{eqnarray}
which can be further approximated as
\begin{eqnarray}
&&\tilde V^{2\textrm{ii}, \,\text{contact}}_{D^+D^0 K^+}+\tilde V^{2\textrm{v}, \,\text{contact}}_{D^+D^0 K^+}+\tilde V^{3, \,\text{contact}}_{D^+D^0 K^+}\simeq \frac{m_D}{f^4}\left(\frac{m_D}{4}-\frac{m_K}{2}\right),
\end{eqnarray}
and considering the mass of the $D$-meson as nearly four times the mass of kaon, becomes
\begin{eqnarray}
&&\tilde V^{2 \textrm{ii}, \,\text{contact}}_{D^+D^0 K^+}+\tilde V^{2 \textrm{v}, \,\text{contact}}_{D^+D^0 K^+}+\tilde V^{3, \,\text{contact}}_{D^+D^0 K^+}\simeq \frac{m_D m_K}{2f^4}.
\end{eqnarray}
It is interesting to see that this result is same as the one obtained in the previous subsection (see Eq.~(\ref{app})), where different three-body contact interactions are obtained from the Lagrangian based on the SU(4) symmetry, and there is a three-body contact interaction coming directly from the Lagrangian. 

Once again, we can compare the sum of the three-body interactions with the contribution of the on-shell two-body $t$-matrices to the three-body diagrams, which is 
\begin{align}
&\tilde V^{2 \textrm {ii,\,on}}_{D^+D^0 K^+}+\tilde V^{2\textrm{v,\,on}}_{D^+D^0 K^+} = \frac{1}{64 f^4}\left(\frac{1}{\left(K_2^\prime + K_3^\prime - K_1\right)^2 - m_K^2}\right)\nonumber\\
&\times \biggl[4\left(m_K^2 + K_3^\prime\left(K_2 - K_3 \right) + 3 K_2 K_3\right)\left(m_D^2 + K_1\left(K_2^\prime - K_1^\prime \right) + 3 K_1^\prime K_2^\prime\right)\biggr.\nonumber\\
&+ 4\left(m_K^2 + K_3 \left(K_2^\prime - K_3^\prime \right) + 3 K_2^\prime K_3^\prime\right)\left(m_D^2 + K_1^\prime\left(K_2 - K_1 \right) + 3 K_1 K_2 \right)\biggl.\biggr].\label{Hon}
\end{align}
Following the same  procedure, as done in the previous subsection, we use Eqs.~(\ref{approx1})-(\ref{approx4}) and deduce Eq.~(\ref{Hon}), up to the first order in $\Delta E$, $\Delta E^\prime$, to get
\begin{align}
&\tilde V^{2 \textrm {ii,\,on}}_{D^+D^0 K^+}+\tilde V^{2\textrm{v,\,on}}_{D^+D^0 K^+} =\frac{1}{f^4}\frac{\left(2 m_D + m_K \right)}{4 m_K^2 \Delta E + 8 m_K mD \Delta E  -4 m_K^2 \Delta E^\prime}  \nonumber\\
&\times \Biggl( -2 m_D^3 m_K + \frac{\Delta E^\prime m_D^2 m_K\left(3 m_D + 2 m_K\right)}{( 2 m_D + m_K)} + \Delta E m_D^2 \left(2 m_D -  m_K\right)\Biggr).
\end{align}
Using the same values for the variables $\Delta E$, $\Delta E^\prime$ and the mass of the kaon, we obtain
\begin{eqnarray}
&\tilde V^{2 \textrm {ii,\,on}}_{D^+D^0 K^+}+\tilde V^{2\textrm{v,\,on}}_{D^+D^0 K^+} \simeq - \dfrac{18.9 m_D^2}{f^4}, \label{Honval}
\end{eqnarray}
whereas Eq.~(\ref{Hoff}) gives
\begin{eqnarray}
&&\tilde V^{2 \textrm{ii},\,\text{contact}}_{D^+D^0 K^+}+\tilde V^{2 \textrm{v},\,\text{contact}}_{D^+D^0 K^+}+\tilde V^{3,\,\text{contact}}_{D^+D^0 K^+}\simeq \frac{0.12 m_D^2}{f^4}.\label{Hoffval}
\end{eqnarray}
It is interesting to notice that the results obtained in Eqs.~(\ref{su4valon}), (\ref{su4valoff}) and Eqs.~(\ref{Honval}), (\ref{Hoffval}) are strikingly similar. We can, thus, conclude that the three-body contact interactions arising from different sources, such as those from the off-shell parts of the two-body $t$-matrices and those coming directly from the Lagrangian, cancel each other in the limit of massless light quarks. In a realistic case, they lead to a small contribution, implying that the results obtained by taking the on-shell parts of the $t$-matrices to calculate three-body interactions should be reliable. 
\clearpage
\bibliographystyle{apsrev4-1}
\bibliography{/Users/amartine/Documents/papers/Bibliography/refs}
\end{document}